\documentclass[letterpaper]{easychair}

\usepackage{graphicx}
\graphicspath{ {images/} }
\usepackage{caption}
\begin{document}

\titlerunning{}

\authorrunning{}

\begin{titlepage}
	\centering
%	\includegraphics[width=0.15\textwidth]{example-image-1x1}\par\vspace{1cm}
%	{\scshape\LARGE Concordia University \par}
%	\vspace{1cm}
%	{\scshape\Large project Report \par}
%	\vspace{1.5cm}
	{\huge\bfseries Return Oriented Programming - Exploit Implementation using functions\par}
	\vspace{2cm}
	{\Large Sunil Kumar Sathyanarayan \\ \small sunilsathyanarayan@protonmail.com\\  \large \vspace{0.5cm} And \\ Katayoun Aliyari \\  \small kt.aliyari@gmail.com  \par}
	\vfill
	supervised by\par
	Dr.~Makan Pourzandi \\ \small pourzand@ciise.concordia.ca
	
	\vfill
	
	% Bottom of the page
	{\large \today\par}
\end{titlepage}

\setcounter{page}{1}
\pagenumbering{arabic}
%\newpage
\section{Abstract}
Return Oriented programming was surfaced first a decade ago, and was built to overcome the buffer exploit defence mechanisms like ASLR, DEP (or W\^\ X) by method of reusing the system code in the form of “gadgets” which are stitched together to make a Turing complete attack. And to perform Turing complete attack would require greater efforts which are quite complex, and there is very little research available for performing a Turing complete attack. So, in this project, we are systemising the knowledge of the existing research that can be used to perform a Turing complete ROP attack.    
\section{Introduction}
Return –Oriented –Programming (ROP) is a technique by which an attacker can induce arbitrary behavior in a program by diverting the program control flow, without injecting any code. A return-oriented program chains together short instruction sequences already present in a program’s address space, each of them ends in a “return” instruction. There are different ways to demonstrate ROP exploitation, one popular demonstration is to deactivate ASLR which is short form of Address Space Layout Randomization and it is a common defense for ROP attacks which works by randomly moving the segments of a program around in memory, preventing the attacker from predicting the address of useful gadgets. So, deactivating ASLR at the beginning of the implementation is a common demonstration. In this work, for our own simplicity, we tried to implement an attack on ROP by deactivating ASLR before implementing the exploit and writing our program. After ASLR deactivation, we implement our program in which a buffer overflow would occur by defining a function in our program which points to another function called function 1 and it also points to function 2 and so on which we explain more in section 5. 
\section{Related Work}
In order to find a way to exploit an attack in ROP, we went through some solutions against the attack such as G-free which is the only general solution as each defense method, rest of the mechanisms provide solutions to a specific built.  We also found different methods to attack ROP such as 'Automated ROP', 'ROP without return' and ‘Return-to-libc’ which we will summarize in the following subsections. In addition, we found various tools to find gadgets for performing ROP namely 'ROPgadget' and'ROPeme' %todo
\subsection{G-Free}
As \cite{gfree}, G-free is a compiler based approach against any possible form of ROP. It can eliminate all the unaligned free-branch instructions which are located inside an executable binary and protect them against attacker data misuse. This method provide an executable gadget-free solution which removes links between necessary chain sequence instructions that cannot be targeted by any possible ROP attack.  In this solution, the first step is to eliminate any unaligned free-brunch instructions. The second step is to protect aligned free-brunch instructions to be secure against misuse.  To achieve this goal, \cite{gfree} employs two techniques: ret instructions to encrypt return address and a cookie-based technique to protect jmp*/call*. Ret instructions provides a short header and saved the encrypted return addresses into the stack. By this technique, whenever the attacker jump into a position of a function, he reaches the footer.  As a result, these jumps transfer the attacker to an incorrect address that attacker cannot control. 
\subsection{Tools for finding gadgets }
In return oriented programming, the core idea is to get useful instruction sequences from the code and chain these instructions together. For this, the attacker should collect some useful sequence of instructions and then reuse these sequences as basic blocks to execute the code. The most important factor in ROP is that these collections of code provide a set of functionalities which allow the attacker to achieve touring completeness without any code injecting. In the next step, the attacker should chain these code sequences in an order to manipulate the program control flow. Gadgets are these valid sequences of instructions to change the control flow satisfyingly \cite{gfree}. In this section, we introduce some existing tools which enable us to find gadgets and chain them together.
\section{Background}
ROP was built to overcome the shortcomings of Buffer overflow, where attacker was able to insert his arbitrary code in the stack segment and execute it, which was prevented by making the stack segment not executable and introducing ASLR made it more difficult for injection of malicious code. Hence ROP's were built exploiting the existing defense mechanisms and reusing the code.  
\subsection{Bufferoverflow}
In a buffer overflow attack, the attacker tries to overflow the stack by exceeding the limited length of the stack and modify the return address and points it to the address of his injected malicious code [4]. According to [7], buffer overflow attack exploits lack of array or buffer bound in compiler. In figure 1, a typical stack layout had been shown in which when a function had been called, how some stack entries like return address can be corrupted by a malicious copy operation.
\break

\begin{center}
%\centering
\includegraphics[scale=1.4]{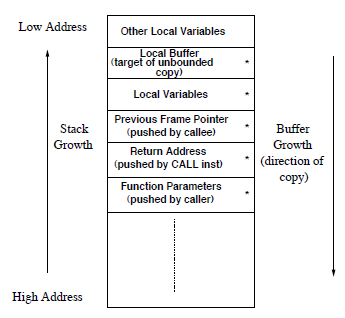}
\begingroup
\captionof{figure}{Stack showing bufferoverflow working}
\label{fig1}
\endgroup
\end{center}
%\hfill \break
\section{Our Work}
Aim of this section is to demonstrate chaining of gadgets, so we wrote a simple buffer exploit from where we chained various types of functions viz., Functions with or without Parameters, and chaining of global functions and their respective behavior on stack which is the replication of actual gadgets in a general way but there are much of assembly coding involved while using gadgets. For this, we have referred blog posts published by various researchers\cite{dhaval}\cite{arcana}\cite{exploitdb}. 

For the Implementation we used a 32-bit version of Debian based Linux Mint 18.1 operating system, and our exploit code was written in C program compiled with GCC complier, gdb for debugging, and extensively used objdump for finding addresses of chaining functions. Most modern Operating systems implement the randomizing the Address layout (ASLR) which makes difficult for a buffer exploit to execute since the code segment of stack is randomized preventing the attacker obtaining the return address of the functions, but for a 32 bit system attackers have often find a way to brute force.
But for our work we considered disabling the ASLR manually during the compilation time
\begin{lstlisting}
gcc vuln.c -o vuln -fno-stack-protector
\end{lstlisting}
-fno-stack-protector disables the stack smashing protector thereby the addresses of the program are not randomized by the compiler, which enables us to control the code sequence.

On the complied code we run \textit{objdump –d vuln2} to obtain the exact memory addresses of the functions and to know the buffer length, and exploit the code sequence by overflowing the buffer and modifying the return addresses to arbitrary code snippets.  
%\begin{lstlisting}
%	objdump –d vuln2
%\end{lstlisting}
Code Snippet below shows exploit code where we have induced a buffer overflow in the function echo() from where we will be executing arbitrary function and variables to demonstrate the chaining of gadgets in two phases: first phase with functions without argument and Second phase is functions with arguments and global variables.
\newpage
%\lstinputlisting[language=C]{vuln.c}
\begin{lstlisting}
#include<stdio.h>
#include<stdlib.h>
//global variable
char str[20] = "MyROPExploit";  
//function without parameters    
void SecretFunctionWithoutParm(){
printf("Welcome to secret function without parameters"); 
}
//function with parameters 
void SecretFunctionWithParm(char argv[]){
printf("Welcome to another secret function with parameter %s\n",argv); 
}
//function with buffer exploit
void echo(){
char buffer[20];
printf("Enter some text:\n");
scanf("%s", buffer);			//buffer overflow 
printf("You entered: %s\n", buffer);  
}
int main(){
echo();
return 0;
}
\end{lstlisting}

In order to chain gadgets, we should learn about two important things about finding addresses of the functions and the length of the buffer. Both of them can be learnt to use either gdb debugger or objdump. we demonstrate using objdump.
\subsection{Finding length of buffer and return address}
\begin{center}
	%\centering
	\includegraphics[scale=0.5]{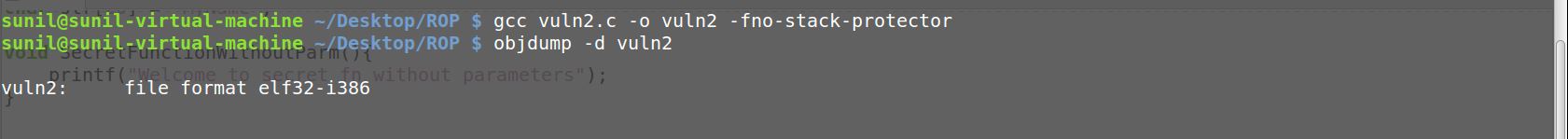}
	\includegraphics[scale=0.5]{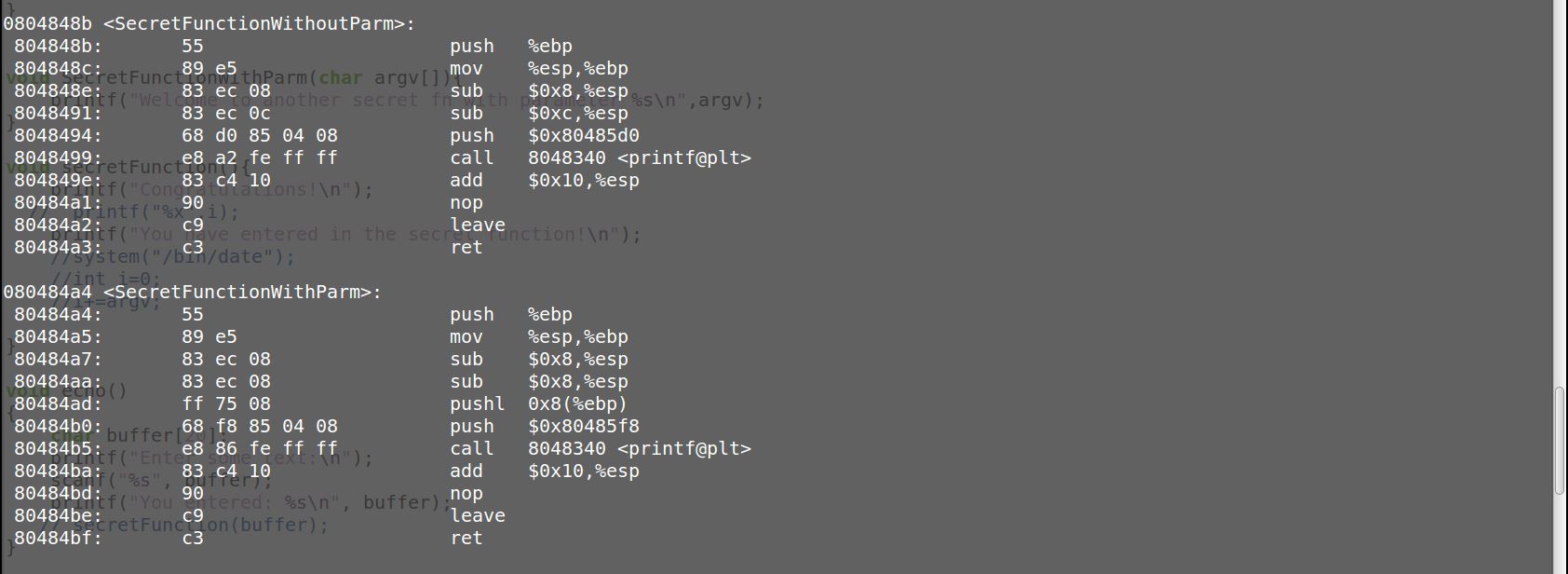}
	\includegraphics[scale=0.5]{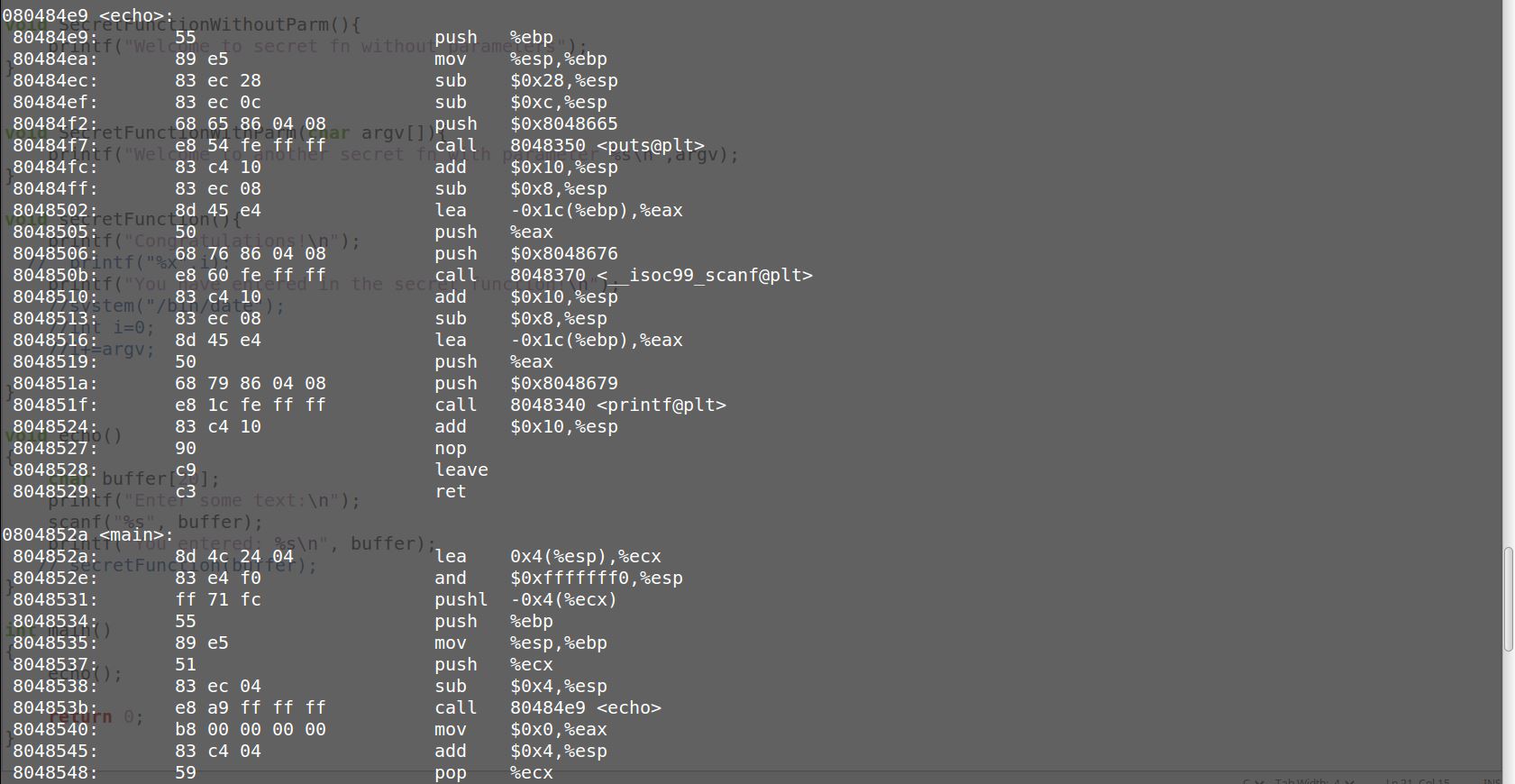}
	\begingroup
	\captionof{figure}{object dump of exploit program}
	\label{fig2}
	\endgroup
\end{center}
 figure 2 shows the snippet of objdump for the exploit code, from echo() function we can see the compiler allocation for the buff variable 
 \begin{lstlisting}
 8048502:	8d 45 e4             	lea    -0x1c(%ebp),%eax
 \end{lstlisting}
compiler has allocated 0X1C or 28 bits for variable so in order to overflow we need to use a buffer of length 32 bits to achieve the bufferoverflow.
   \begin{center}
   	%\centering
   	\includegraphics[scale=0.7]{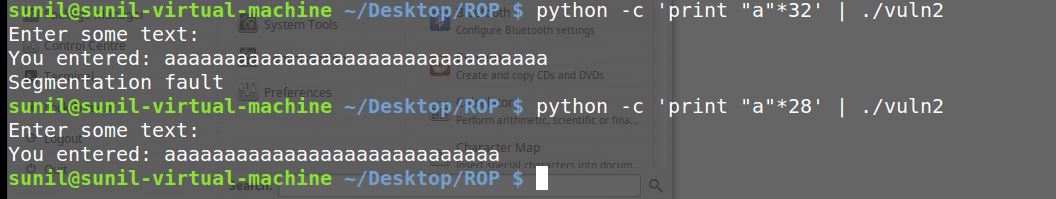}
   	\begingroup
   	\captionof{figure}{bufferoverflow}
   	\label{fig3}
   	\endgroup
   \end{center}
To find the address of functions, we can just use the starting addresses provided by objdump
\begin{lstlisting}
080484a4 <SecretFunctionWithParm>:
0804848b <SecretFunctionWithoutParm>:
\end{lstlisting} 
\subsection{ROP Exploit with and without arguments}
As we explained in previous subsection we have address of functions required for the exploit.Now we have to replace the original return address, but to know where to insert return address First we should understand how the stack in figure 4 works.
  \begin{center}
  	%\centering
  	\includegraphics[scale=0.55]{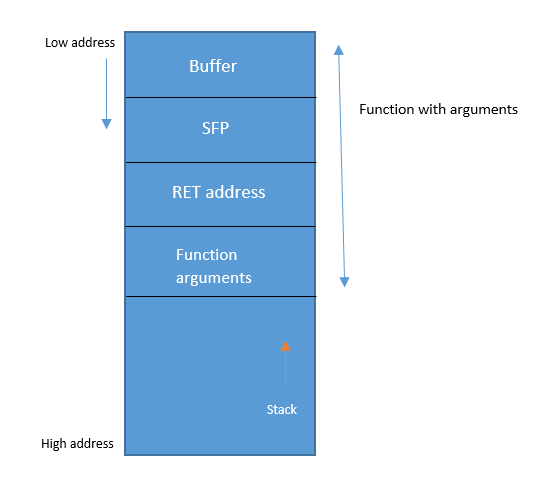}
  	\begingroup
  	\captionof{figure}{stack overview of a normal function with paramaters}
  	\label{fig4}
  	\endgroup
  \end{center}
 Function is stored on stack starting from a high memory address to and it grows towards lower addresses and stores RET address of invoking function in our case main function.
 and the value of buffer variable is stored from a lower address and grows towards higher memory address. By overflowing the values of buffer variable we can modify the actual return address of the invoking function to some arbitrary code segment and execute malicious action as shown in figure 5.
 \begin{center}
 	%\centering
 	\includegraphics[scale=0.55]{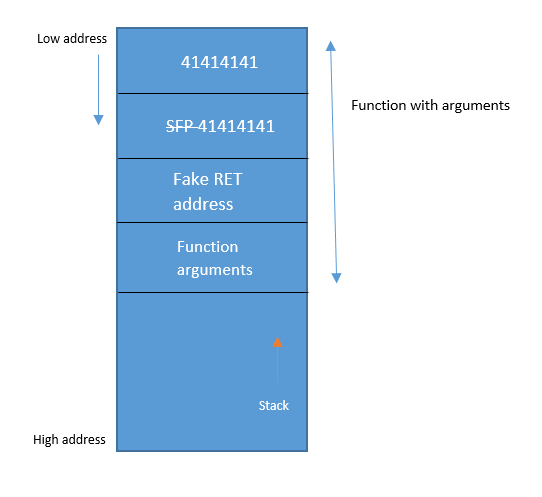}
 	\begingroup
 	\captionof{figure}{stack overview of a normal function with fake return address}
 	\label{fig5}
 	\endgroup
 \end{center} 
  Now we use this concept and inject the address of SecretFunctionWithParm 080484a4 to our exploit and obtained the intended results as in figure 6.
  \begin{center}
  	%\centering
  	\includegraphics[scale=0.6]{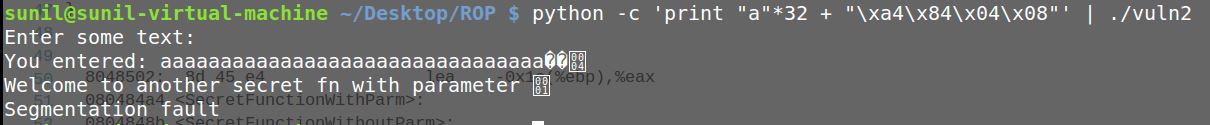}
  	\begingroup
  	\captionof{figure}{Demo of injecting fake return address to SecretFunctionWithParm()}
  	\label{fig6}
  	\endgroup
  \end{center} 
  Now since we are able to insert a arbitrary function, we try inserting arbitrary arguments to the exploit functions in figure 7
    \begin{center}
    	%\centering
    	\includegraphics[scale=0.95]{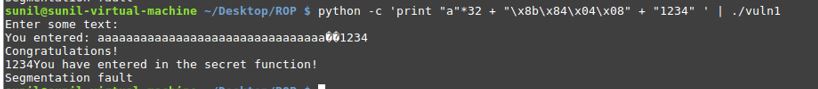}
    	\begingroup
    	\captionof{figure}{Demo of injecting fake return address with parameter to SecretFunctionWithParm()}
    	\label{fig7}
    	\endgroup
    \end{center}
    And now to demonstrate the chaining of gadgets, by chaining the  SecretFunction() iteratively in figure 8 
        \begin{center}
        	%\centering
        	\includegraphics[scale=0.6]{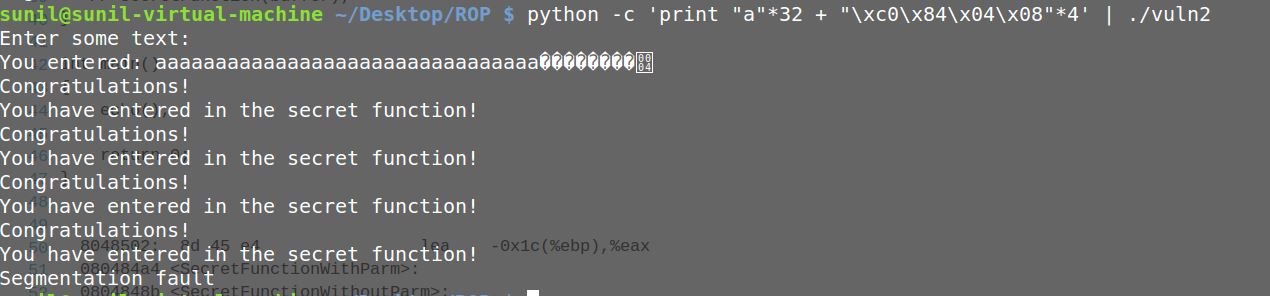}
        	\begingroup
        	\captionof{figure}{Demo of injecting fake return address to SecretFunction() iteratively}
        	\label{fig8}
        	\endgroup
        \end{center}
      And lastly we want show inserting of global variable to SecretFunctionWithParm() for this we found the address of global variable using gdb debugger and results can be seen on figure 9
      \begin{center}
      	%\centering
      	\includegraphics[scale=0.51]{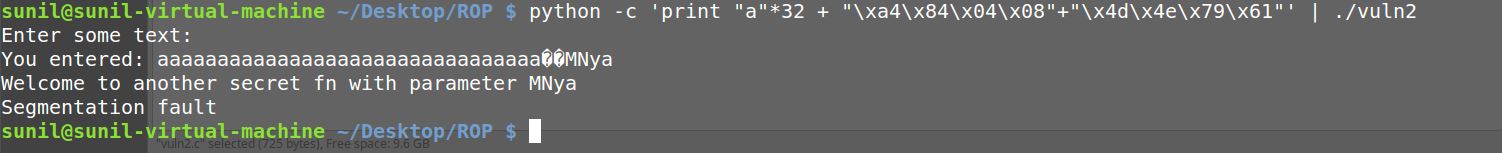}
      	\begingroup
      	\captionof{figure}{Demo of injecting global variable to SecretFunctionWithParm()}
      	\label{fig9}
      	\endgroup
      \end{center}
      
%\section{Experimental results}
\section{Conclusion} 
Return Oriented Programming maybe a decade old and not many exploits were reported using ROP, and it is stealth feature, which cannot be detected by the Intrusion Detection Systems or any Signature based detection systems as it reuses the system trusted library to perform the malicious action.
Many Defense mechanisms have been proposed like G-Free\cite{gfree} which can provide a solution to all the systems, but most of other works focused on providing system specific solutions, and implementation of G-free is not widely adopted, Hence Research on ROP seems to be very limited so in our project we demonstrated gadget chaining in form of chaining functions which is exact replication that can be used to perform a turing complete attack.

\clearpage

%\section{Appendices}

%\subsection{Appendix A - Report Contributions of team members}

%\begin{table}[h!]
%	\centering
%	\begin{tabular}{| l | p{10cm} |} 
%		\hline
%		Name & Sections\\ [0.5ex] 
%		\hline\hline
%		Sunil Kumar Sathyanarayan & \begin{itemize} \item Section 1 \item Section 5 \item Section 6  \end{itemize}\\
%		\hline
%		Katayoun & \begin{itemize} \item Section 2 \item Section 3 \item Section 4  \end{itemize}\\
%		\hline
	
%	\end{tabular}
%	\caption{Report Contributions} 
%	\label{tab:tb_team_contributions}
	
%\end{table}


\begin{thebibliography}{99.}%

%\bibitem{quest_pwd} J Bonneau, C Herley, P.C Oorschot, F. Stajano (2012) A Framework for Comparative Evaluation of Web Authentication Schemes, \url{http://www.cl.cam.ac.uk/techreports/UCAM-CL-TR-817.pdf}
\bibitem{gfree} Kaan Onarlioglu, Leyla Bilge, Andrea Lanzi, Davide Balzarotti, and Engin Kirda. 2010. G-Free: defeating return-oriented programming through gadget-less binaries. In Proceedings of the 26th Annual Computer Security Applications Conference (ACSAC '10). ACM, New York, NY, USA, 49-58. DOI=10.1145/1920261.1920269 -\url{http://doi.acm.org/10.1145/1920261.1920269} 

\bibitem{kalilinux} Kali Linux - \url{https://www.kali.org/}

\bibitem{dhaval} Dhaval Kapil buffer-over-flow-exploit - \url{https://dhavalkapil.com/blogs/Buffer-Overflow-Exploit/}
\bibitem{arcana} Introduction to return oriented programming (ROP) by Alex Reece - \url{http://codearcana.com/posts/2013/05/28/introduction-to-return-oriented-programming-rop.html}
\bibitem{exploitdb} Return-Oriented-Programming
(ROP FTW) by By Saif El-Sherei
 - \url{https://www.exploit-db.com/docs/28479.pdf}\\
%\bibitem{kalilinux} Kali Linux - \url{https://www.kali.org/}

\end{thebibliography}
\end{document}